# Elusive Threats: Security Weaknesses of Commercial Cellular Networks

*Scott C. Forbes, Ph.D.*

*"...abandoning the public interest can be caused by many things, of which timidity and desire for personal security are the most insidious."*
— Howard Morgan, to President Kennedy in 1963[1]

## I. Understanding Policy Making in the Telecommunications Industry

Morgan's tone closely resembles that of several modern researchers who argue that policy makers with the influence to increase telecommunications access levels in the United States—primarily those in the FCC—have been negligent in their duties. One reason for this negligence includes their preoccupation with maintaining their government positions. But their timidity in policy formulation has deeper roots, namely a lack of necessary industry and technical knowledge and their generally complacent attitude toward enhancing public access to information networks.

One such researcher arguing for a more active approach to telecommunications access policy making, Jorge Schement, co-director of the Institute for Information Policy at the Pennsylvania State University, spoke eloquently of the dangers of a complete lack of regulation during the academic forum held by the FCC during the drafting of its 1999 Strategic Plan: "with complete de-regulation, you get lord of the flies; not chaos, but the imposition of rules on behavior by a very small number." He then argued that access should serve as a core value for the FCC, something that is not currently the case:

> "Focusing on access it seems to me is the ultimate justification for any mission of an organization like the FCC in a democracy because short of that, it's not anchored to anything that we hold dear in terms of our basic values. It's anchored basically to behavior and at the point that the behavior is either resolved or the players don't chose to play, that justification goes away."

Schement goes on to suggest that access is a local concept, and that the FCC's federalist view of access is now detrimental because "we [academic researchers] increasingly know that local circumstances shape the opportunities for access all across the nation."[2] Schement then argues that the goals of the FCC should not reside in competition or access per se, but rather in social, economic, and political participation. To the extent that access policies can promote such participation, they should be supported.[3] To this end, Schement offers several suggestions for enhancing the effectiveness of telecommunications access policies:

---

[1] Howard Morgan, letter to the President, January 23, 1963, Washington DC. Quoted in (Kohlmeier 1969), p. 82.
[2] Schement is referring to work, including his own, that has shown numerous circumstances across the United States where local circumstance override national access trends and assumptions. See [Schement, 1995 #241; Schement, 1996 #243; Schement, 1997 #245; Schement, 1998 #246; Schement, 1998 #247; Schement, 1998 #396; Schement, 1999 #248; Schement, 1999 #249; Schement, 1999 #250; Schement, 1999 #251; Schement, 1999 #252; Schement, 2000 #253; Mueller, 1994 #194; Mueller, 1996 #196; Mueller, 1997 #197].
[3] (Commission ) p. 37-40

1) Recognize the role maintenance costs play in long-term access levels, regardless of the service in question:

> "So if we're going to subsidize, we shouldn't subsidize access. Access isn't the problem. We should subsidize maintenance. Helping people stay on once they're on is much more likely to keep people on and raise telephone rates from where they -- penetration rates from where they have been than if we were focusing on access."[4]

2) Increase demographic data collection and dissemination by the FCC:

> "The FCC ought to be looking to share with others; ought to be looking to build the kind of data-collecting regime that allows others to participate in collecting the data, but also in sharing it.[5]

3) Promote functional access, or helping people acquire the access they want, not just access that Congress or other government bodies think people should or should not have:

> "Americans make choices on how to access and we ought to facilitate those choices because what we want to do is facilitate access. It shouldn't matter to us how they access. It should matter to us that they access in functional ways and that functional is a social construction that they decide for themselves."[6]

Acting on these or any other suggestions for substantive policy change in the telecommunications sector is easier said than done. William E. Kennard, Chairman of the FCC, commented on the difficulty of affecting change in telecommunications access policy in the United States:

> "…one of the most important lessons that I've learned is how difficult it is in this town, given the way policy is made in Washington today, to stay focused on your agenda and keep it moving, because the whole infrastructure of the town is designed to co-opt your agenda. It is an unbelievable thing."[7]

Kennard discussed one instance—his FCC confirmation—where his ability to generate and regulate telecommunications policies was affected by outside interests groups:[8]

> You know, we are daily assaulted by people who are paid a lot of money to basically take the agenda away from the five commissioners. And it's the most amazing thing. I mean,

---

[4] (Commission ) p. 123
[5] (Commission ) p. 85-86
[6] (Commission ) p. 124
[7] (Commission 1999b) p. 82
[8] An interest groups is an intermediary between citizens and government. Dozens of books and hundreds of articles exist on defining and elaborating interest groups and how they affect governmental decision making. The definition listed above is both to narrow the term and thus increase its usefulness, at least for my purposes. More lengthy definitions and discussions can be found, in, among others, (Berry 1977), (Mahood 2000), (Olson 1965), (Schattschneider 1961), and (Stone 1996).

> I've never had a job where so many people are paid so much money to talk to me and tell me what to do.
>
> And it's not just the, you know, the people who are paid to do their work, the lawyers and the lobbyists, but it's the think tanks, it's the PR campaigns. It's the way that people are able to get hearings called in Congress.
>
> And you're called up there and not only are the hearings instigated often in this way, but the questions are planted. The most amazing thing happened to me in my confirmation. I was -- the night before my confirmation, one kind soul faxed me a stack of questions that were going to be asked of me in my confirmation.
>
> Many of them had been drafted by industry groups and not only -- they had the letterheads right there, you know, ask him this; if he says this, say this. You know, it's an amazing thing. And so we are challenged here, and I know many of you are in government, so you face the same challenges to find the true public in the public interest because what we hear most of, what is so easy for us to do is to hear what the shareholders' interests are, the economic interests, the industry interests. And sometimes, when you're in my position, just trying to articulate what this means to the average consumer or people who rely on these services is sort of an aside or an afterthought, because the press is always -- they're here... [9]

Both Schement and Kennard illustrate well the complex telecommunications policy creation process and the difficulty in not only creating a viable new policy, but keeping that policy from being watered down or buried by opposing policy makers. Especially difficult is promoting telecommunications policy, defined by Aufderheide as "calculated government intervention in the structure of businesses that offer communications and media services"[10], that does more than support a Market-oriented viewpoint. A brief discussion why it is difficult to promote a non-Market viewpoint, particularly at the FCC, is given below.

As noted in Chapter 3, the FCC historically took two views of access regulation, broadcasting and the common carrier. Both views were grounded in the organizational structure of the telecommunications industry in the early 1900s and were viewed by the FCC as successful because they promoted a smooth working of the corporations within these sectors of the economy. With these views and an organizational structure built around distinct industry segments, the FCC has been ill-prepared to react to emerging new technologies and political pressures advocating change.[11]

---

[9] (Commission 1999b) p. 82-83

[10] (Aufderheide 1999) page 5. She considered the intervention calculated because all policy decisions inherently involve risk—the risk involved in deciding between alternatives with unknown outcomes. (Dahl 1989) page 75. (Bernstein 1998) on p. 8 makes an interesting observation that the word "risk" is derived from the Italian word "risicare" which means "to dare"—viewing the policy-making process as a challenge among parties instead of a cooperative experience fits more with the perception of corporate decision making process than "objective" policy making by government bodies.

[11] (Fried 1999; Huber 1997; Hundt 1997; Powell 1998a)

While the FCC remained content to operate under these two regulatory paradigms, in the late 1980s and early 1990s, it came under increasing pressure from Congress and already dominant telecommunications industry lobbyists to move toward a Market-centric outlook, primarily because the Internet and other new technologies, notably wireless, cable, and broadband telecommunications access, gave existing companies the opportunity to expand their reach into the consumer's home. Unable to combat this political pressure, the FCC Commissioners, already closely-associated with the corporations they themselves ostensibly regulated, shifted their regulatory ideology to more closely align themselves with their economist staffers.

This ideological shift resulted in the FCC's nearly wholesale endorsement of the Market as the most effective tool for "regulating" the telecommunications industry. Policies challenging the legitimacy or validity of the Market's ability to distribute licenses, promote access, or promote competition were viewed with much skepticism. The main objective of the FCC became not the regulation of the telecommunications industry, but rather the promotion of competition in the telecommunications industry.

The FCC saw a competitive marketplace as a means of self-regulation for the corporations and an efficient distributive mechanism for telecommunications goods and services, such as access to the telephone. FCC reports beginning in the mid 1990s showcased this new emphasis on promoting the Market above all else. And perhaps unwittingly, the FCC, by moving towards this view, effectively reduced its legitimate role as a regulatory body, and lessened the need for its existence.[12]

In sum, the FCC has become an organization whose primary mission is to promote competition in telecommunications industry, with all other policies viewed as helping agents for the Market. Therefore, access policies are promoted only if they support a more competitive marketplace. The public interest, once seen as counteracting market failures, is now viewed by the FCC as synonymous with competition, and thus has been co-opted and relegated to irrelevancy.[13]

Thus the FCC makes the decisions it does because it now views itself as a the promoter of competition instead of as an active regulator. Or, to look at it in another way, the FCC sees as the best way to regulate the American telecommunications industry the promotion of free market policies. As a result, it has taken the position that while the Market does not immediately solve all issues of inequity, it has the potential to do so in the future and to counteract the forces of the Marketplace prematurely could negatively affect the telecommunications industry more than it would aid it.

---

[12] (Albiniak 1999; Chen 2000a; Chen 2000b; Commission 1997b; Commission 1999a; Crowe 2000b; FCC 1999; Kennard 1998; Kennard 1999; Kumar 2000; Stone 2000)

[13] (Commission 1995; Commission 1996; Commission 1997a; Commission 1998; Furchtgott-Roth 1998; Kennard 1999; Krasnow 1998; Lathen 1999a; Lathen 1999b; Lundstedt 1977; Mueller 1997; Ness 1998; Permut 1998; Powell 1998b; Rowland 1997)

As it relates to promoting telecommunications access policy in the public interest, the FCC is unlikely to change its current policies of promoting marketplace competition above all else. With bi-partisan Congressional pressure firmly leading the FCC to deregulate as opposed to regulate, funding may become an issue if the FCC does not comply with this pro-Market directive. Furthermore, as mentioned earlier, the Commissioners are already beholden to some degree to the industry for post-Commission careers, and to not promote pro-Market policies that allow for expansion could be dangerous for their future job prospects. Additionally, most FCC staffers now employed take a decidedly pro-Market approach, and thus internal FCC reports already support a pro-Market viewpoint. There is simply no entity within the FCC's immediate vision disagreeing with the pro-competitive viewpoint, especially since the Public and non-corporate interest groups are given little input into the FCC's decision-making process.[14]

What is most disturbing about the FCC and its role in telecommunications policy is the small likelihood that its stance will change in the near future. As its funding and legitimacy is in many way determined by Congressional support and industry acknowledgement of its regulatory power, there is no current incentive for the FCC to alter its focus of promoting competition to the virtual exclusion of all other policies. Furthermore, since this ideology has been prevalent within the FCC for at least five years, to change it would be a time consuming process, and would have to be replaced with another, equally powerful viewpoint.

Additionally, since both access and the public interest have been successfully consumed by the reigning "deregulate and favor the Market" viewpoint, even these two policies are not seen as competitors to the incumbent Market-driven one. In a way, the true insidiousness of the FCC's pro-Market focus is not the Commission's false assumption that the Market can solve anything. It is instead the assumption that all policies, regardless of whether they explicitly associate themselves with the Market, are indeed best served by the Market, and should thus be left to the Market. In short, all policies not explicitly incorporating the Market into their structure are seen as being unaware that the Market can indeed solve the issues mentioned in the policy, and thus need to have the FCC explicitly associate them with the Market, such as the current issue of broadband access.[15]

To change the dominant pro-Market paradigm in telecommunications regulation as articulated by the FCC would require three things: 1) The introduction of FCC staff and Commissioners who oppose the laissez-faire view of the Marketplace in favor of a more realistic view that market failures occur on a regular basis and that government-led policy changes, such as policies promoting telecommunications access, can rectify such failures;

---

[14] (Commission 1997b; Commission 1999a; Commission 1999b; Commission 2000)
[15] (Albiniak 1999; Bloomer 1999; Culver 2000; Dickson 2000; Lathen 1999a; Magazine 2000; Now 1999; Ozer 1999; Speta 2000)

2) The removal or lessening of the political pressures by Congress, corporate lobbyists, and other institutions to focus solely on promoting competition to the exclusion of other legitimate policies, such as telecommunications access; 3) The introduction of a new telecommunications access policy theory that offers a practical means for incorporating a carefully articulated definition of the public interest into the policy creation process. As the first two deal with altering the regulatory structure and are best done in person, this dissertation will instead strive to obtain the third item.

Before the new telecommunications policy access theory is described, however, it makes sense to first describe the telecommunications industry in more detail and to build on Chapter 3's discussion of how the FCC has neglected promoting access and the public interest in favor of the Marketplace and how such actions are detrimental to consumers and the Public at large.

Two examples in particular highlight how current telecommunications policies, especially those promoted by the FCC, have not only failed to secure a more competitive marketplace, but have hindered telecommunications access in the public interest in the meantime. These examples, the local and long distance telephony market and the changing characteristics of the average telecommunications user, are discussed below. Following this discussion, a theory promoting telecommunications access in the public will be introduced that shows how such actions can be avoided in the future.

A. Local and Long Distance Telephony

The term telephony, once meaning only voice communications, is increasingly blurred. "Telephone" companies now routinely send great volumes of data across their once "voice-only" networks; the four remaining RBOCs (also known as ILECs)[16] saw at least a 25% gain in their data revenue in 1999, compared to 1998, and expect even larger gains in 2000.[17] See Tables 4.1-4.3 for specific 1999 revenue and percentage amounts.

Table 4.1    Total Revenue of Bell Companies 1999[18]

| Company | Revenue (in billions) |
|---|---|
| SBC | $49.5 |
| Bell Atlantic | $33.2 |
| BellSouth | $25.2 |
| US West | $13.2 |

---

[16] Regional Bell Operating Companies—the local telephone companies of the Bell System prior to the breakup of AT&T in 1984. Bell Atlantic and BellSouth are two examples. RBOCs are also referred to as ILECs, or incumbent local exchange carriers. The competitor carriers are termed CLECs, or competitive local exchange carriers.
[17] (Cantwell 2000)
[18] Source for Tables 4.1-4.3 is (Cantwell 2000).

Table 4.2    Data as Percentage of Total Bell Company Revenue 1999

| Company | Percentage |
| --- | --- |
| US West | 12.9% |
| SBC | 11.5% |
| BellSouth | 11.1% |
| Bell Atlantic | 8.7% |

Table 4.3    Wireless as Percentage of Total Bell Company Revenue 1999

| Company | Percentage |
| --- | --- |
| SBC | 13.7% |
| Bell Atlantic | 13.7% |
| BellSouth | 12.7% |
| US West | 1.8% |

Also changing, although much more slowly, is the degree of competition in local and long distance markets. Despite the FCC's rhetoric, the Telecommunications Act of 1996 has not promoted significant competition, and competition has not drastically affected either the local or long distance markets. For instance, RBOCs still dominate as providers of virtually all the major metropolitan areas in the United States, serve close to 75% of the population[19], and according to the FCC's 2000 report on competition in the telecommunications industry bill 96% of 1998 local service revenues.[20]
The vast majority of new growth and most of the new competition in the telecommunications industry has come in the form of competitive local exchange carriers (CLECs) that compete with the Bell companies for customers in local markets. The total number of US facilities-based local competitors has doubled since 1998's 160 to 333 by the end of 1999. Total access lines have increased almost as much: from 5.5 million in 1998 to 10.4 in 1999 and as access lines have increased, so too has revenue from competitive local switched services: $3,500 million in 1998 to $6,300 million in 1999.[21]
The RBOCs have not welcomed these new competitors; they have in many cases have thwarted the upstarts' attempt to gain and retain customers: The RBOCs hamper even simple installations of telecommunications services, purposefully process inaccurate CLEC, and delay billing and provisioning schedules.[22] Keep in mind that these tactics come years after the passage of the 1996 Act forbidding such practices and have for the most part gone unpunished by the FCC. In fact, the FCC has even supported incumbent RBOCs in their desire to enter other markets, despite the fact that they continue to hinder competitors in their own local markets: the FCC approved Bell Atlantic's entry into the

---

[19] (Commission 1995) p. 46
[20] (Commission 2000) p. 47
[21] (Wilson 2000)
[22] (Geppert 1999)

New York long distance market despite the Department of Justice's recommendation to reject the application or at least approve with more stringent conditions.[23]

With Bell Atlantic's entry, and other RBOCs as well—US West recently won approval to provide long distance service in Texas—the long distance sector is beginning to see competition. Much like the local market, however, entrenched players still dominate the long distance market and dictate the access prices and services available to consumers. For instance, average retail prices for interstate long-distance services in the US have dropped only 5% since the 1996 Act was passed.[24]

An analysis of the market structure of the industry using a standard economic metric, the Herfindahl-Hirschmann Index (HHI), further illustrates the relatively little progress made by "Market-focused" policies of the FCC. The HHI is a summary measure of the market structure of an industry; it is a measure of industry concentration and by extension monopoly power. The HHI has been a staple of industrial organization economics for thirty years and is held in high regard by policy makers, regulators, and other legislators since it takes into account a market's absolute numbers of firms, their relative sizes, and gives greater weight to dominant firms with high market share.[25]

The HHI is calculated as the sum of the squared market shares of each company in the industry. Pure competition, with an infinite number of competing firms, would result in an HHI of 0, while a market dominated exclusively by one company—a monopoly—would be 10,000, since one firm would have 100% of the market share: $(100)^2 = 10,000$. As an example, Table 4.4 lists the long distance telephony market's HHI as 8,155 in 1984: $(90.1)^2 + (4.5)^2 + (2.7)^2 + (2.6)^2 = 8,155$.

According to the Horizontal Merger Guidelines issued by the U.S. Department of Justice and the Federal Trade Commission, a market with an HHI value under 1,000 is considered unconcentrated and mergers within such an industry that leaves the HHI under 1,000 is not seen as adverse to competition. A market with an HHI value between 1,000 and 1,800 is considered moderately concentrated and merger increases that raise the HHI more than 100 points "raise significant competitive concerns," according to the guidelines. Markets with an HHI over 1,800 are considered highly concentrated and any

---

[23] (Crowe 2000a). Section 271 of the 1996 Act requires that four criteria be met before a long-distance application by an RBOC will be granted. First, the RBOC must show that one local competitor is using its network and providing local service via an interconnection agreement or if no local competitor is using the network, the RBOC must provide a statement by the state commission showing the RBOC's willingness to offer the use of its network. Second, nondiscriminatory access must be established with the a detailed 14 point checklist that includes specific access and service level requirements. Thirdly, the RBOC must not aid its long distance affiliates in gaining unfair advantages such as using existing customer information to promote new broadband services. Finally, the applicant must demonstrate that its entry into the LD market would promote the public interest.

[24] (Poor's 1999)

[25] See (Hirschmann 1964) for an early explanation of the index. See (Greer 1992) for a more recent discussion with more current journal references, Chapter 7.

mergers that raise the HHI by more than 50 points in such an industry raises antitrust concerns.[26] In short, markets with HHI's under 1,000 are considered competitive, those between 1,000 and 1,800 are moderately competitive, and those over 1,800 likely to be dominated by a monopoly or oligopoly

AT&T was a virtual monopoly at the time of its breakup in 1984, as shown in Table 4.4. While competition in the long distance market has improved since then, recent market consolidation efforts threaten to reverse the trend of the last 15 years; the recent MCI—WorldCom increased the HHI for the first time since the breakup of AT&T in 1984 and did so in an already highly concentrated industry. In fact, according to the DOJ's own guidelines, the merger should not have taken place, yet the FCC approved it. And AT&T still has a dominant position in many parts of the country: when measured at the state level, AT&T's market share is never less than 40% and rises to 75% in some states.[27]

---

[26]http://www.usdoj.gov/atr/hhi.htm
http://barney.sbe.csuhayward.edu/~skamath/powerpoint/econ3551s7/sld001.htm
http://www.minneapolisfed.org/pubs/fedgaz/00-01/HHI.html

The following is the relevant excerpt the U.S. Department of Justice and the Federal Trade Commission's Horizontal Merger Guidelines, found at: http://www.usdoj.gov/atr/public/guidelines/horiz_book/toc.html:

1.51 General Standards

In evaluating horizontal mergers, the Agency will consider both the post-merger market concentration and the increase in concentration resulting from the merger.(18) Market concentration is a useful indicator of the likely potential competitive effect of a merger. The general standards for horizontal mergers are as follows:

a) Post-Merger HHI Below 1000. The Agency regards markets in this region to be unconcentrated. Mergers resulting in unconcentrated markets are unlikely to have adverse competitive effects and ordinarily require no further analysis.

b) Post-Merger HHI Between 1000 and 1800. The Agency regards markets in this region to be moderately concentrated. Mergers producing an increase in the HHI of less than 100 points in moderately concentrated markets post-merger are unlikely to have adverse competitive consequences and ordinarily require no further analysis. Mergers producing an increase in the HHI of more than 100 points in moderately concentrated markets post-merger potentially raise significant competitive concerns depending on the factors set forth in Sections 2-5 of the Guidelines.

c) Post-Merger HHI Above 1800. The Agency regards markets in this region to be highly concentrated. Mergers producing an increase in the HHI of less than 50 points, even in highly concentrated markets post-merger, are unlikely to have adverse competitive consequences and ordinarily require no further analysis. Mergers producing an increase in the HHI of more than 50 points in highly concentrated markets post-merger potentially raise significant competitive concerns, depending on the factors set forth in Sections 2-5 of the Guidelines. Where the post-merger HHI exceeds 1800, it will be presumed that mergers producing an increase in the HHI of more than 100 points are likely to create or enhance market power or facilitate its exercise. The presumption may be overcome by a showing that factors set forth in Sections 2-5 of the Guidelines make it unlikely that the merger will create or enhance market power or facilitate its exercise, in light of market concentration and market shares.

[27] (Commission 2000) p. 84

Combine this trend with the rapid consolidation of the local telephone market as shown in Figure 4.6[28] and it becomes apparent that more competition has not been a result of the 1996 Act or the FCC regulation of the telecommunications industry[29], and that serious access issues—particularly those in the public interest—still remain.[30]

Table 4.4  Market Share of US Long Distance Carriers 1984-1998[31]

| YEAR | AT&T | MCI WorldCom | | Sprint | All Other Long Distance Carriers | Herfindahl-Hirschman Index (HHI) * |
| | | MCI | WorldCom | | | |
| --- | --- | --- | --- | --- | --- | --- |
| 1984 | 90.1 % | 4.5 % | | 2.7 % | 2.6 % | 8,155 |
| 1985 | 86.3 | 5.5 | | 2.6 | 5.6 | 7,479 |
| 1986 | 81.9 | 7.6 | | 4.3 | 6.3 | 6,783 |
| 1987 | 78.6 | 8.8 | | 5.8 | 6.8 | 6,298 |
| 1988 | 74.6 | 10.3 | | 7.2 | 8.0 | 5,720 |
| 1989 | 67.5 | 12.1 | 0.2 % | 8.4 | 11.8 | 4,778 |
| 1990 | 65.0 | 14.2 | 0.3 | 9.7 | 10.8 | 4,527 |
| 1991 | 63.2 | 15.2 | 0.5 | 9.9 | 11.3 | 4,321 |
| 1992 | 60.8 | 16.7 | 1.4 | 9.7 | 11.5 | 4,074 |
| 1993 | 58.1 | 17.8 | 1.9 | 10.0 | 12.3 | 3,795 |
| 1994 | 55.2 | 17.4 | 3.3 | 10.1 | 14.0 | 3,466 |
| 1995 | 51.8 | 19.7 | 4.9 | 9.8 | 13.8 | 3,197 |
| 1996 | 47.9 | 20.0 | 5.5 | 9.7 | 17.0 | 2,823 |
| 1997 | 44.0 | 19.0 | 6.5 | 9.5 | 20.9 | 2,431 |
| 1998 | 43.1 | 25.6 | | 10.5 | 20.9 | 2,641 |

---

[28] Figure 4.3 does not include all of AT&T's activities since its inception as the parent company of the Bell System in 1899. Instead, it emphasizes changes in AT&T's corporate structure following its 1984 divestiture. Therefore, more emphasis is placed on the Regional Bell Operating Companies than such notable components of AT&T as Western Electric and Bell Labs.

[29] The HHI, while still the dominant indicator of monopoly power, is not foolproof. The first drawback is that it is often difficult to define what market is under consideration, especially in the telecommunications industry. Secondly, in industries where a limited number of players actively engage in vigorous price wars, abuse of monopoly/duopoly power is often questionable, such as in the soft-drink industry where Coca-Cola and PepsiCo control 75% of the market and arguably in the long-distance market when referring to calling plan packages. Newer econometric methods that focus instead on price increases as a result of mergers are now used in conjunction with the HHI (Staff 1998).

[30] Since 1998, there has been further consolidation in the local market. Only four of the original RBOCs now exist, and only Bell South has not merged with either a previous RBOC or a long-distance carrier. Furthermore, the remaining local companies are still virtual monopolies in their respective areas. To be fair, there has also been an increase in local competitors and the RBOCs are slowly moving into the long distance market. Yet as these companies vie for customers, and as cable and wireless networks merge with traditional local and long distance networks, the large players will account for relatively more revenue than they do today, thereby further increasing their market power.

[31] Table 4.4 and 4.5 shows total market share based on revenue (Commission 2000).

Table 4.5　　　Residential Market Share of US Long Distance Carriers 1995-1998

|  | AT&T | MCI WorldCom* | Sprint | Teleglobe (Excel)** | Other |
|---|---|---|---|---|---|
| **Access Lines*** | | | | | |
| 1995 | 74.6 % | 13.0 % | 4.2 % | N.A | 8.3 % |
| 1996 | 69.9 | 14.1 | 5.0 | 2.8 % | 8.2 |
| 1997 | 67.2 | 13.2 | 5.7 | 3.8 | 10.1 |
| 1998 | 62.6 | 15.1 | 5.7 | 3.6 | 13.0 |
| **Toll Revenue** | | | | | |
| 1995 | 68.5 % | 14.6 % | 5.6 % | N.A. | 11.3 % |
| 1996 | 63.3 | 16.0 | 6.6 | N.A. | 14.1 |
| 1997 | 61.1 | 16.6 | 5.6 | 3.8 % | 13.0 |
| 1998 | 58.3 | 18.4 | 5.7 | 3.3 | 14.3 |
| **Direct Dial Minutes** | | | | | |
| 1995 | 69.5 % | 16.1 % | 5.8 % | N.A | 8.6 % |
| 1996 | 62.5 | 15.9 | 7.1 | 3.1 % | 11.4 |
| 1997 | 62.4 | 14.9 | 6.5 | 3.7 | 12.5 |
| 1998 | 58.4 | 17.0 | 6.5 | 3.7 | 14.3 |

Figure 4.6  Evolution of AT&T 1899-2000

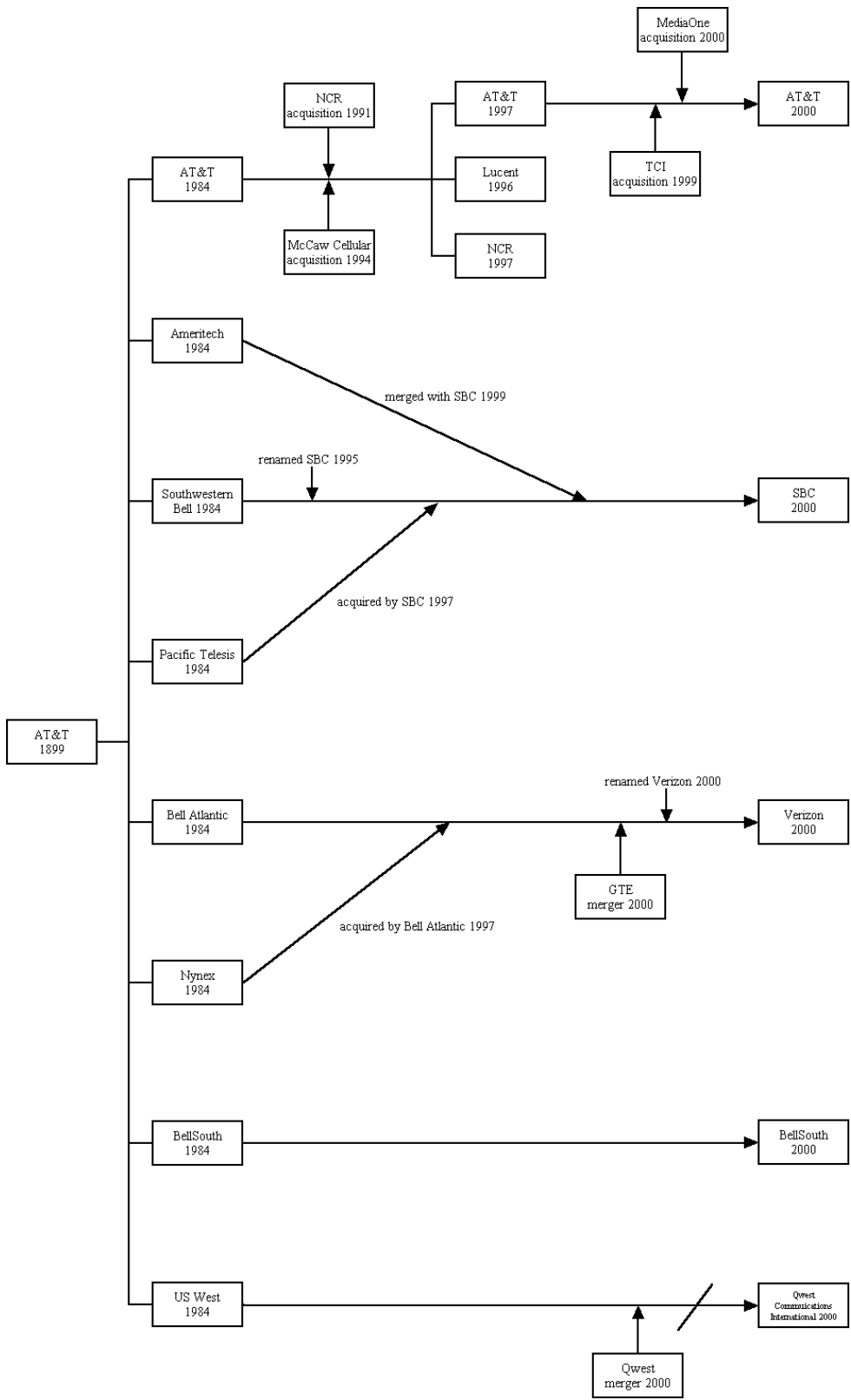

B.     Internet Services

Whereas the FCC has investigated at a high level Internet service and its associated issues of open access, broadband access, and the Digital Divide, it has not formulated specific policies to address Internet access at an individual or household level.  Instead, it has allowed the Market to dictate penetration rates and levels.  For many households, this has worked well.  Yet access policies in the public interest address more than just the wealthiest or the earliest adopters.  They address those who for whatever reason are unable to acquire and retain access to such services without assistance, whether that assistance be private or government-based.

Since private corporations in the United States tend only to sell products and services and ignore the broader social implications of their actions, it often falls on the government to address these social implications.  Yet as has been said earlier in this chapter and in Chapter 3, the FCC now believes that the Market should address these implications.  The practical implications of such a viewpoint are that the corporations become tantamount to the Market and anything they do is seen as legitimate and effective market forces.  And, as just said, the corporations are more concerned with the profit-oriented consequences of their actions and less so with the less tangible effects upon the Public.   This is unfortunate because as seen below Internet access is becoming a dominant form of Public interaction and participation and must be addressed by specific telecommunications access policies.

While the number of users and their and annual expenditures on telephony products has remained relatively stable, as shown in Table 4.7, the number of Internet users has grown drastically (Table 4.8), as has their diversity.  While household income is still positively correlated with Internet use,[32]  the fastest growing segment of new Internet users are those with annual household incomes between $25,000 and $49,00 per year.  In fact, by 2003, households earning under $35,000 will comprise one third of the estimated 162 million home users in the US, and they are expected to fuel Internet usage beyond its current household penetration level of 50%.[33]

Table 4.7     Annual Household Telephony Expenditures 1995-1998[34]

---

[32] (Census 1999) Table 922
[33] (Trager 2000)
[34] (Commission 2000) p. 26  Table is in current dollars.

|      | Local Exchange Carriers | | Long Distance Carriers | | Wireless Carriers | | Total Expenditures | |
|------|---------|--------|---------|--------|---------|--------|---------|--------|
|      | Average | Median | Average | Median | Average | Median | Average | Median |
| 1995 | $358    | $305   | $250    | $136   | $62     | $0     | $670    | $515   |
| 1996 | 359     | 309    | 250     | 132    | 83      | 0      | 692     | 531    |
| 1997 | 379     | 323    | 305     | 183    | 100     | 0      | 784     | 610    |
| 1998 | 398     | 336    | 270     | 150    | 119     | 0      | 787     | 616    |

Source: Calculated by IAD staff with data provided by PNR and Associates Inc., *TLC MarketShare Monitor*.

Annual expenditures are based on monthly household bills for those households with wireline telephone service.

The sample does not include households from Alaska and Hawaii.

Table 4.8    US Adult Internet Users 1997-1999[35]

| Period | Number of Users (in millions) |
|---|---|
| Mid 1997 | 52.6 |
| Mid 1998 | 65.1 |
| Year End 1998 | 83.6 |
| Mid 1999 | 100.7 |
| Year End 1999 | 106.3 |

The degree to which the Internet and its related services have affected telecommunications users has also increased drastically. A 2000 survey by *Wired* magazine divided its respondents into three categories: very wired, somewhat wired, and not wired. Those who were very wired used four or more of the following technologies: Internet, cellular or wireless phone, computer, fax, email, online banking or investing, and online shopping. Of note is that the original 1997 survey which inspired this 2000 survey used as its criteria for "very wired" only laptop, cell phone, beeper, and home computer, plus a requirement for email use three times a week. In the 1997 survey, only 9% of the respondents, even by that diminished criteria, were very wired or somewhat wired. Today, that percentage, with the more rigorous definition, stands at 31% for very wired and 46% for somewhat wired, for a total of 77%—an eightfold increase. When asked in the 2000 survey which forms of technology they used most, the 100 year-old technology still prevailed: the telephone (albeit wireless or cellular) with 45%. A close second was the computer (43%), followed by email (37%) and the Internet (31%). When asked "In 10 years, will Internet access become as common as the telephone in US households?", 88% said "Yes."[36] Also according to *Wired*, both "average" and "high-volume" households use email frequently, along with numerous other communication technologies which may be Internet-centric. As shown in Table 4.9, the average household sends/receives 16.4 electronic communications per day while the high-volume user sends/receives 34.3 communications per day with pagers and emails accounting for the greatest difference between the two groups.

Table 4.9    Weekly Communications Sent and Received 2000[37]

| Technology | Average Household | High-Volume Household |
|---|---|---|
| Phone | 54 | 102 |
| Mail | 35 | 50 |
| Email | 16 | 59 |
| Voicemail | 4 | 12 |

---

[35] (Infoworld 2000)
[36] (Breslau 2000)
[37] (Staff 2000)

| Cell phone | 3 | 8 |
| Pager | 2 | 7 |
| Fax | 1 | 2 |
| **Total** | **115** | **240** |

The longer Internet users maintain access, the greater amount of time they spend on the Internet, as shown in Table 4.10. Furthermore, Internet users also purchase more—significantly more—the longer they are online, as shown in Table 4.11. What is also of note is that the effect of online tenure is constant regardless of the goods and services purchased. Whether it is a computer system, a CD, or groceries, the longer someone has been online, the more likely they are to purchase an item.

What this data indicates is that once exposed, online access is viewed by its users as a critical and if not necessary component of their lives. This characteristic of the Internet is reminiscent of the telephone—once people have access to it, they tend not to relinquish it unless under extreme hardship. Together, these tables describing online access show that telecommunications access is expanding and growing; people are not simply moving from the telephone to the Internet; they are maintaining their use of traditional technologies such as the telephone, and increasing their use of the online technologies.

Table 4.10     Hours Per Week Spent Online By Online Tenure 1999[38]

| Group | Hours |
| --- | --- |
| Newbies | 5.4 |
| Intermediates | 6.5 |
| Veterans | 8.2 |

Table 4.11    Online Sales Between Thanksgiving and New Year's Day 1999[39]

| Category | Research | Convenience | Replenish |
| --- | --- | --- | --- |
| % bought by Veterans | 74% | 66% | 77% |
| % bought by Intermediates | 23% | 28% | 23% |
| % bought by Newbies | 3% | 7% | 0% |

---

[38] Newbies are those who have been online under 1 year. Intermediates have been online 1-2 years, and veterans have been online over 2 years (Jupiter 1999).
[39] Researched goods consist of items that a consumer would do background research on before making a purchase; they tend to be expensive items such as computer systems and furniture. Convenience Goods include what marketers have traditionally labeled impulse buys, or those that people buy because they saw or heard about the item. They include books, CDs, and other smaller-ticket items. Replenishment Goods are goods oriented toward basic human necessities, such as groceries and drugs (Inc. 1999).

## II. Comparing Four Contemporary Network Access Theories

"Theory" is a vague word. In communications research, it has several connotations, distinguished by McQuail (1994) into four types: social scientific, normative, operational, and everyday.[40] Social scientific theories focus on observation and hypothesis testing, and may incorporate prediction of agent behavior. Normative theories focus on social values and help show how media should operate or behave and why. Operational theories are practical in nature; they show how practical "industry knowledge" is applied by its practitioners. Everyday theory is concerned primarily with describing the individual knowledge people have based on their own experiences.

The Network Access Theory (NAT) as introduced in Chapter 1 is a broad normative theory taking a macro approach that advocates a change in the way policy is created in the United States, particularly as it relates to telecommunications access in the public interest. Being normative, the theory proposes a view of how policy should be created, who the interested parties should be, what objectives should be strived for, and what the outcomes should be in a successful telecommunications access policy. Succinctly, the goal of the NAT is to explicitly include the public interest in the telecommunications access policy creation process. The purpose of this inclusion is to increase the Public's social, economic, and political participation via the creation and implementation of proactive and progressive policies at the national and local levels.

While the NAT is normative, it does have some social scientific qualities, specifically in its use of quantitative data and reliance upon a comparative analysis of competing theories based upon pre-established criteria. These non-normative qualities help to give it credence and move beyond its marking as a "wish list" theory—the NAT can be effectively implemented in the "real world."

A. Competing Access Theories
Three telecommunications access policies based on other access theories will be used for comparison purposes to highlight the strengths of NAT-based policies relative to their competitors. The assumption behind this policy analysis is that theories provide the general worldview and main assumptions upon which policies are built. These policies are the practical manifestations of the theories, and can be more closely and accurately evaluated than the more general theories, especially a normative theory. Furthermore, policies usually have more direct impact on the environment in which they are introduced.

For clarity, a NAT-based policy is hereafter referred to as a Network Access Policy (NAP). To clarify, the NAT is the foundation upon which a NAP is created. A NAP is therefore the product of the NAT. As such, the measure of success for the NAT is a policy—a NAP—that incorporates all of the NAT's required qualities and successfully achieves the theory's objectives, namely the promotion (and tangible increase) in Public access to the Nation's telecommunications network. In short, a successful policy is an indicator of a successful theory, so a successful NAP indicates that the NAT is a successful theory. The presence of this equivalency is why the a NAP will be compared with the policies of other theories to demonstrate the NAT's superiority as a means for restructuring the telecommunications access policy creation process.

---
[40] (McQuail 1994) p. 4-6

The criteria used to judge the policies is based on the core values and structure of the NAT. Perhaps not surprisingly, the NAP excels in each one of these areas and overall is "rated" higher than its three competing policies. The intent of the comparison was not to set up a "straw man" argument whereby the three competing policies were weak from the start and the NAP was overpowering. This would have resulted in an illogical argument—the NAT is the better theory because the NAP derived from it more thoroughly represents the characteristics of the NAT. This would also have been a waste of time since the theory as stated is already normative, in that it takes as one of its core assumptions that it intends to promote specific values that other theories neglect. In that sense, there is no need for such an illogical justification for the NAT's legitimacy.

Instead, the analysis is to demonstrate that the value of the NAT resides in several of its unique characteristics. The NAT has the ability to structure a telecommunications policy creation capable of generating polices which accurately reflect the Public's interests regarding network access. Additionally, no other access policies have such capabilities. This is not a surprising argument; if a policy existed and had been implemented that could already meet the objectives of the NAT, then the promotion of telecommunications access in the public interest would have already been achieved.

So, the analysis of the different policies should be regarded as an illustration of the different perspectives, assumptions, and objectives of the parent theories. For instance, the analysis will demonstrate how the NAT places more emphasis than the other theories on several issues, particularly: 1) Diverse yet concerted policy maker participation, 2) Practical implementation of viable technologies to facilitate access, and 3) Promotion of telecommunications access that promotes social, political, and economic participation.

The three competing theories are: Free Market Theory, FCC Access Theory, and ICM Universal Service Theory. Since the alternative theories are not the primary focus of this dissertation, they will not be reviewed in as much detail as the NAT has been. What follows is a brief summary of the alternative theories followed by an evaluation of the policies each theory generates.

The Free Market Theory (MKT) is defined by its reliance upon capitalistic market forces as articulated by orthodox economists—supply, demand, profit motives, individual freedom, and corporate accumulation of wealth—to self-correct inefficiencies within the telecommunications sector of the economy. As such, government policies are not needed to correct these inefficiencies; they will only hinder what the market is already capable of doing.[41] Examples of efficiencies that the Market is supposed to be able to correct include: unequal access to goods and services despite a consumer's ability to pay, prices for goods higher than people are willing or able to pay, and unfair competition within a marketplace.

FCC Access Theory (FCC) is based upon the assumption that all people in the United States should have reasonable access to the nation's telecommunications network at an affordable price. While this definition is moving toward including Internet-based services and products, it is still rooted in historical universal service policies that rely upon the telephone as the technology of choice as well as a common carrier view of regulation. Recently FCC Access theory has begun to borrow assumptions and objectives found in the Free Market Theory.

ICM Universal Service Theory (ICM) builds upon the FCC Access Theory to explicitly include advanced telecommunications services such as the Internet as well as several elements lacking in the FCC policy including individual choice, a recognition of local versus national circumstances, and the belief that access policies should be proactive, not reactive.

The ICM was underpinned by four interconnected principles: availability, connectivity, affordability, and choice.[42] The four elements were inseparable; without any piece of the triangle, the entire structure falls. This tight integration is what gave the ICM of universal service its potency: any mandate meeting the ICM's stringent requirements presumably had overcome the difficulties at each stage of development; its "bugs" have been worked out.

Finally, there is a realization in the ICM that the Market cannot self-correct all market inequities and that corporate assistance could be of use to government regulators in their pursuit of increased Public telecommunications access. This realization is perhaps the most distinct difference between the ICM, FMT, and FCC theories and is the closest link to the NAT.

---

[41] Chapter 3 discusses this viewpoint in more detail, especially when it references FCC reverence to the Market.

[42] See the following references for a formulation of a version of an Internet-based universal service program, the Informed Choice Model (ICM). The pyramid structure was taken from this body of work as well (Schement 1995; Schement 1998; Schement 1999; Schement 2000).